\documentclass[prl,showpacs,twocolumn,amsmath]{revtex4}

\usepackage{latexsym}
\usepackage{epsfig}
\usepackage{graphics}
\usepackage{amsmath}
\usepackage{amssymb}
\usepackage{color}

\newcommand{\be}{\begin{equation}}
\newcommand{\ee}{\end{equation}}
\newcommand{\bea}{\begin{eqnarray}}
\newcommand{\eea}{\end{eqnarray}}
\newcommand{\nr}{\nonumber\\}

\begin{document}

\title{Full Potential Multiple Scattering for X-ray Spectroscopies}

\author{Keisuke Hatada$^{1,2}$}
\author{Kuniko Hayakawa$^{3,2}$}
\author{Maurizio Benfatto$^2$}
\author{Calogero R. Natoli$^2$}
\affiliation{$^1$Dipartimento di Fisica ``E. Amaldi",
 Universit\`a degli studi Roma Tre, 
 Via Vasca Navale 84, Rome, I-00146 Italy,\\
$^2$Laboratori Nazionali di Frascati - INFN, Via E. Fermi 40,
    00044 Frascati, Italy,\\
$^3$Museo Storico della Fisica e Centro Studi e Ricerche ``Enrico Fermi",
    Via Panisperna 89A, 00184, Roma, Italy}

\date{\today}

\begin{abstract}

  We present a Full Potential Multiple Scattering (FP-MS) scheme for the interpretation of several X-ray spectroscopies that is a straightforward generalization of the more conventional Muffin-Tin (MT) version. Like this latter, it preserves the intuitive description of the physical process under consideration and overcomes some of the limitations of the existing FP-MS codes. It hinges on a fast and efficient method for solving the single cell scattering problem that avoids the convergence drawbacks of the angular momentum (AM) expansion of the cell shape function and relies on an alternative derivation of the multiple scattering equations (MSE) that allows us to work reliably with only one truncation parameter, {\it i.e.} the number of local basis functions in the expansion of the global scattering function determined by the classical relation $l_{\rm max} \sim k \, R$. 

\end{abstract}

\pacs{61.05.js,61.05.cj,71.15.Ap}
\keywords      {Multiple scattering, non-muffin-tin potential,
                X-ray absorption, X-ray Photoelectron-Diffraction}

\maketitle

%

Multiple scattering theory ( MST ) has been widely used to solve the
Schr\"odinger equation (SE) ( or the associated Lippmann-Schwinger equation (LSE)) both for scattering and bound states. It was proposed originally by Korringa and by Kohn and Rostoker (KKR) as a convenient way to calculate the electronic structure of solids ~\cite{korringa47,kohn54} with potentials of the muffin-tin (MT) form ({\it i.e.} potentials that are bounded by non overlapping spheres and spherically symmetric), and was later extended to polyatomic molecules by Slater and Johnson ~\cite{slater72}. However the MT approximation cannot properly describe a great number of physical systems, ranging from open lattices to molecular systems with substantial anisotropy ( {\it e.g.} systems of biological interest ), to surfaces and interfaces.

The attempts to extend MST to space-filling potential cells in order to eliminate the interstitial region and take full account of the asphericity of the potential, have generated a lot of controversies that have gone on for more than twenty years ~\cite{butler92}. Many (but not all) of the questions are now settled and we refer the reader to the book of Gonis and Butler ~\cite{gonis00} for a comprehensive review of the state of the art in this field. Usually the applications of the space-filling method have regarded mainly the calculations of the electronic structure of solids, {\it i.e.} states below the Fermi level. Applications to states well above the Fermi energy, as required in the simulations of x-ray spectroscopies, like absorption, photo-emission, anomalous scattering, etc..., have been scarce, probably due to the difficulties encountered in the numerical implementation of the method. We mention here the work by Huhne and Ebert ~\cite{huhne99} on the calculation of
 x-ray absorption spectra using the full-potential spin-polarized relativistic MST,
that of Ankudinov and Rehr ~\cite{Ankudinov05} 
and that of Foulis ~\cite{foulis95} based on a version of the MST that uses spherical cells and treats the interstitial potential in the Born approximation~\cite{natoli90,foulis90}. 

All these methods, however, have their limitations and drawbacks. The method used by Foulis treats in an approximate way the potential in the interstitial region and moreover looses one of the major advantages of the MST, namely the separation between dynamics and geometry in the solution of the scattering problem. Huhne and Ankudinov 
use the potential shape function to generate the local basis functions which are at the heart of MST. The expansion of the shape function and the cell potential in spherical harmonics leads to a high number of spherical components in the coupled radial equations that becomes progressively cumbersome to handle and time consuming with increasing energy and in absence of symmetry. This feature might also be at the origin of another problem related with the saturation of "internal" sums in the MSE ~\cite{gonis00}, as discussed below. Moreover no critical discussion is devoted in their work to the convergence problems of MST.
 
The purpose of this letter is to give an alternative derivation and interpretation of the FP-MS equations that will allow us to work with square matrices for the phase functions $S_{ LL' }$ and $E_{ LL' }$ and for the cell $T_{ LL' }$ matrix (see below) with only one truncation parameter, contrary to the present accepted view ~\cite{gonis00}. As a result this scheme can be viewed as a natural extension of its MT counterpart, with all the consequent advantages for the interpretation of X-ray spectroscopies. In connection with this we shall present a new scheme to generate local basis functions for the truncated potential cells that is simple, fast, efficient, valid for any shape of the cell and reduces to the minimum the number of spherical harmonics in the expansion of the scattering wave function. 

In order to solve for scattering states we seek a solution of the SE continuous in the whole space with its first derivatives, satisfying the asymptotic boundary condition 
$
\psi({\bf r};{\bf k}) \, \simeq \, \left( \frac{k}{16\pi^3} \right)^{\frac{1}{2}} \, \left[ {\rm e}^{{\rm i}
{\bf k} \cdot {\bf r}} + f(\hat{\bf r};{\bf k})
 \frac{{\rm e}^{{\rm i} {k r}}}{r} \right],
$
\noindent where ${\bf k}$ is the photo-electron wave-vector.
We partition the space in terms of non overlapping space-filling cells $\Omega_j$ with surfaces $S_j$ and centers at ${\bf R}_j$. This is equivalent to partitioning the overall space potential $V({\bf r})$ into cell potentials, such that $V({\bf r}) = \sum_j v_j({\bf r}_j)$, where $v_j({\bf r}_j)$ takes the value of $V({\bf r})$ for ${\bf r}$ inside cell $j$ and vanishes elsewhere. Here and in the following ${\bf r}_j = {\bf r} - {\bf R}_j$. The partition is assumed to satisfy the requirement that the shortest inter-cell vector ${\bf R}_{ij} = {\bf R}_i - {\bf R}_j$ joining the origins of the nearest neighbors cells $i$ and $j$, is larger than any intra-cell vector ${\bf r}_i$ or ${\bf r}_j$, when ${\bf r}$ is inside either cell $i$ or cell $j$.
We also assume that there exists a finite neighborhood around the origin of each cell lying in the domain of the cell ~\cite{butler92}. We then start from the following identity involving surface integrals in ${d\hat {\bf r}} \equiv d\sigma$
\bea
\displaystyle{
\sum_{j = 1}^{N} \, \int_{S_{j}} \, \left[ G_{0}^{+}({\bf r}' - {\bf r}) 
{\bf \nabla} \psi({\bf r}) - 
\psi({\bf r}) {\bf \nabla} G_{0}^{+}({\bf r}' - {\bf r}) 
\right] \cdot {\bf n}_{j} \, {\rm d} \sigma_{j} } && \nr
\displaystyle{= \int_{S_{o}} \, \left[ G_{0}^{+}({\bf r}' - {\bf r}) 
{\bf \nabla} \psi({\bf r}) - 
\psi({\bf r}) {\bf \nabla} G_{0}^{+}({\bf r}' - {\bf r})
\right] \cdot {\bf n}_{o} \, {\rm d} \sigma_{o} \, }. && \nonumber
\eea
\noindent  Here $G_{0}^{+}({\bf r}' - {\bf r})$ is the free Green's function and $\Omega_o = \sum_j \Omega_j$, with surface $S_o$. This identity is valid for all ${\bf r}'$ lying in the neighborhood of the origin of each cell, since in this case the integrands are continuous with their first derivatives.

The heart of MST is the introduction of the functions $\Phi_{ L } ({ \bf r }_j)$ which inside cell $j$ are local solutions of the SE with potential $v_j({\bf r}_j)$ behaving as $J_{ L }({\bf r}_j)$ for $r_j \rightarrow 0$. They form a complete set of basis functions such that the global scattering wave function can be locally expanded as $\psi ( { \bf r }_j ) = \sum_{ L } A_{ L }^j ({\bf k}) \Phi_{ L } ( { \bf r }_j )$ ~\cite{butler92}.
Using this expansion in the above surface integrals, taking ${\bf r}'$ in the neighborhood of the origin of cell $i$, so that $G_{0}^{+}({\bf r}' - {\bf r})\equiv G_{0}^{+}({\bf r}_i' - {\bf r}_i) = \sum_L J_L( {\bf r}_{i}') \tilde{H}_L^{ + } ( { \bf r }_{i})$ (since ${\bf r}$ is confined to lie on the cell surfaces), and putting to zero the coefficients of $J_L( {\bf r}_{i}')$ due to their linear independence, we readily arrive at the MST compatibility equations for the amplitudes $A^{j}_{ L } ({\bf k})$ (see Ref.~\cite{gonis00} page 129 for analogous derivation in the case of bound states)
\be
\sum_{jL'} H_{L L'}^{i j} A^{j}_{L'}({\bf k}) = Y_{L}(\hat{\bf k}) {\rm e}^{{\rm i} {\bf k} \cdot {\bf R}_i } (k/\pi)^{1/2} = I^{i}_{L}({\bf k}) 
\label{si}
\ee
\noindent where 
$$
 H_{ LL' }^{i j}  =  \int_{ S_j } [ \, \tilde{H}_L^{ + } ( { \bf r }_i ) 
     \nabla \Phi_{ L' } ( { \bf r }_j ) - \Phi_{ L' } ( { \bf r }_j )
     \nabla \tilde{H}_L^{ + } ( { \bf r }_i ) \, ] \cdot { \bf n }_j \, 
     {\rm d}  \sigma_j 
$$
\noindent and the rhs term comes from the outer sphere $\Omega_o$ (see Appendix A of Ref. ~\cite{sebilleau06} for details). As usual $J_{ L }({\bf r}) = j_l ( kr ) Y_{L} ( \hat {\bf r} )$ and $\tilde{H}_{ L }^+ ({\bf r}) = -i k h_l^+ ( kr ) Y_{L} ( \hat {\bf r} )$. The usual derivation of the MSE now proceeds by re-expanding $\tilde{H}_L^{ + } ( { \bf r }_i )$ around center $j$ by use of the equation $\tilde{H}_L^{ + } ( { \bf r }_i ) = \sum_{L'} G^{i j}_{L L'} J_{L'} ( { \bf r }_j )$, where $G^{i j}_{L L'}$ are the free electron propagator in the site and angular momentum basis ( KKR real space structure factors). Unfortunately this relation introduces a further expansion parameter into the theory (with related convergence problems) which is actually unnecessary, as shown below.

We in fact observe that the integrals over the surfaces of the various cells $j$ can be calculated over the surfaces of the corresponding bounding spheres (with radius $R_b^j$) by application of the Green's theorem, since both $\tilde{H}_{ L }^+ ({\bf r})$ and $\Phi_{ L } ( { \bf r })$ satisfy the Helmholtz equation outside the domain of the cell (where the potential is zero). We then use the following relation 
\be
\int_{ S_j } \, Y_{L'} ( \hat {\bf r}_j ) \nabla \, \tilde{H}_L^{ + } ( { \bf r }_i )  \cdot { \bf n }_j \, {\rm d}  \sigma_j = 
G^{i j}_{L L'} \, \frac{\rm d} {{\rm d} R_b^j} \, j_{l'} ( kR_b^j )
\label{main}
\ee
(and the similar one without derivatives) which is exact for all $L$ provided $|{\bf r}_i - {\bf r}_j| = R_{ij} > r_j $ for ${\bf r}$ lying on the surface $S_j$. This is a consequence of the fact that under this condition the series $\tilde{H}_L^{ + } ( { \bf r }_i ) = \sum_{L'} G^{i j}_{L L'} J_{L'} ( { \bf r }_j )$ converges absolutely, since $ G^{i j}_{L L'} J_{L'} ( { \bf r }_j ) \le (r_j/R_{ij})^{l'} 1/(kR_{ij})^{l+1} [2(l'+l)+1]^l $ for fixed $l$, as can be seen by using the usual expression for $G^{i j}_{L L'}$ and the asymptotic expansion of the Bessel functions for high values of the index $l'$ ($l' \gg kr_j$) ~\cite{hatada07}. By use of the Weierstrass criterium, the series is also uniformly convergent in the entire solid angle domain and can therefore be integrated term by term ~\cite{whittaker65}, leading to the desired result (this property is also true for the series derived with respect to ${\bf r}$) .

By inserting in Eq. (\ref{si}) the expression for the basis functions expanded in spherical harmonics $\Phi_{ L } ( { \bf r } ) = \sum_{ L' } R_{ L'L} ( r ) Y_{ L' } ( \hat {\bf r} ) $ and using Eq. (\ref{main}), we finally obtain 
\be
\sum_{L'} E_{L L'}^{i} A^{i}_{L'}({\bf k})\, - \sum_{j, L', L''} ^{j \neq i} G^{i j}_{L L''} S^{j}_{L'' L'} A^{j}_{L'}({\bf k}) =   I^{i}_{L}({\bf k}) 
\label{ceq}
\ee
\noindent defining
$$
E_{LL'} = R_b^2 W[-ikh_l^+,R_{ L L' }]; \;
S_{LL'} = R_b^2 W[j_l,R_{ L L'} ] \; 
$$
where the Wronskians $W[f,g] = fg'-gf'$ are calculated at $R_b$ and reduce to diagonal matrices for MT potentials. We observe that, even if the potential has a step, the wave function and its first derivatives are continuous, so that the AM expansion converges uniformly in $\hat{\bf r}$~\cite{kellog54} and can be thus integrated term by term. Eq. (\ref{ceq}) looks formally similar to the usual MSE. However we notice that the sum over $L''$ runs over the angular momentum components of the basis functions $\Phi_{ L } ( { \bf r } )$ and is not affected by convergence constraints related to the re-expansion of $\tilde{H}_L^{ + } ( { \bf r }_i ) $ around center $j$. 

To find the local solutions of the SE we do not expand the truncated cell potential to avoid AM expansion problems. Dropping the cell index $j$, we write the SE in polar coordinates for the function $P_L ({ \bf r }) = r \Phi_L ({ \bf r })$
\be
\left[ \frac {d^2}{dr^2} + E - v(r,\hat{\bf r}) \right ]P_L (r,\hat{\bf r}) = 
\frac {1}{r^2} \tilde{L}^2 P_L (r,\hat{\bf r})
\label{se}
\ee
\noindent where $\tilde{L}^2$ is the angular momentum operator, whose action on $P_L (r,\hat{\bf r})$ can be calculated as:
\be
  \tilde{L}^2 P_{ L} (r,\hat{\bf r}) \hspace{-1.5mm } \; = \, \hspace{-1.5mm }
    \sum_{L'} l' ( l' + 1) r R_{L' L} ( r ) Y_{L'} ( \hat {\bf r} ).
     \label{pe} 
\ee

Equation (\ref{se}) in the variable $r$ looks like a second order equation with an inhomogeneous term. Accordingly we use Numerov's method to solve it. As is well known, putting $f_{ i, j }^L = P_{ L} (r_i,\hat{\bf r}_j)$ and dropping for simplicity the index $L$, the associated three point recursion relation is
$$
  A_{ i+1, j } f_{ i+1, j }
  - B_{ i, j } f_{ i, j }
  + A_{ i-1, j } f_{ i-1, j }= g_{ i, j }
  - \frac{ h^6 }{ 240 } f_{ i, j }^{\rm vi} \nonumber
$$
where,
\bea
  A_{ i, j } &=& 1 - \frac{ h^2 }{ 12 }v_{ i, j } \nr
  B_{ i, j } &=& 2 + \frac{ 5 h^2 }{ 6 }v_{ i, j } = 12 - 10 A_{ i, j } \nr
  v_{ i, j } &=& v ( r_i, \hat {\bf r}_j ) - E \nr
  g_{ i, j } &=& \frac{ h^2 }{ 12 } [ q_{ i+1, j } +
                 10 q_{ i, j } + q_{ i-1, j } ] \nr  
  q_{ i, j } &=& \frac{ 1 }{ r_i^2 } \sum_{L'} l' ( l' + 1)
     r_i \, R_{L' L} ( r_i )Y_{L'} ( \hat {\bf r}_{ j } ).
    \nonumber 
\eea
Here $i$ is an index of radial mesh and $j$ an index of angular points  on a Lebedev surface grid ~\cite{lebedev75}. Obviously $R_{L' L} ( r_i ) = \sum_{j} w_{ j } P_{ L} (r_i,\hat{\bf r}_j) / r_i \, Y_{L'} ( \hat {\bf r}_{ j } )$, where $ w_{ j }$ is the weight function for angular integration associated with the chosen grid.

Only the inhomogeneous term $ q_{ i+1,j } $ in the recurrence relation, containing the still unknown term $ f_{ i+1,j } $, prevents us to solve the equation by iteration, from the knowledge of $ f_{ i,j} $ and $ f_{ i-1,j} $ at all the angular points. This difficulty is easily overcome by introducing the backward second derivative formula, whereby 
\bea
  g_{ i,j } 
    \sim \frac{ h^2 }{ 12 }
    \left[ 13 q_{ i,j } - 2 q_{ i-1,j } + q_{ i - 2,j } \right] 
    + \frac{ h^5 }{ 12 } q_{i,j}''' - \frac{ h^6 }{ 24 } q_{i,j}^{\rm iv}
  \label{eqn:inhomo}
\eea
\noindent so that at the cost of a small errors $ O ( h^5 )$ only the backward points $ f_{ i,j} $, $ f_{ i-1,j} $ and $ f_{ i-2,j} $ are now involved. The appearance of $q_i'''$, strictly infinite at the step point, does not cause practical problems. 

In this way the three-dimensional discretized equation can be solved along the radial direction for all angles in an onion-like way ~\cite{hatada07}, provided the expansion (\ref{pe}) is performed at each new radial mesh point. We use a log-linear mesh $\rho = \alpha \, r + \beta \, { \ln } \, r $, to reduce numerical errors around the origin and the bounding sphere.~\cite{brastev66}. 
We tested this modified Numerov method against analytically solvable separable model potentials, with and without shape truncation, obtaining very good results~\cite{hatada07}. 

The next delicate point to tackle is how to handle and truncate the various $L$ sums in Eq. (\ref{ceq}). Here only two truncation parameters appear, the number of basis functions and the number of their AM components, corresponding to the indexes $L'$ and $L$ in $R_{LL'}$. (The external index $L$ in Eq. (\ref{si}), coming from the expansion of $G^+_0$, must coincide with the index $L$ of $R_{LL'}$ when calculating $H^{ii}_{LL'} \equiv E^i_{LL'}$, due to the orthogonality of the spherical harmonics). These two indexes are in principle unrelated, although one can speculate that for $l > k \, R_b$ and positive energies the wave-function hardly sees the anisotropy of the potential. This is in practice what happens in our method of generating the local basis functions, so that the two indexes can be cut safely at the same value $l_{\rm max}\sim k \, R_b$ and the matrices $S$ and $E$ can be treated like square matrices. This is also in keeping with the physical fact that the elements of the atomic T-matrix $T_{ll}=(SE^{-1})_{ll}$ tend rapidly to zero for $l > l_{\rm max}$. The need to converge first the internal sum over $L''$ in Eq. (\ref{ceq}), pointed out by various authors, was probably related to the slow rate of convergence in the AM expansion of the basis functions, due to their method of generation.

We are aware that even in the case of "small" overlap of the cell bounding spheres the $L$ truncation procedure is likely to be divergent ~\cite{butler91}. A simple calculation ~\cite{hatada07} shows that $G(R_{ij})_{ll} \, T_{ll}$ goes like $ (2R_b/R_{ij})^{2l}/l^3$ for $l >> l_{\rm max}$, which explains why for "small" overlap the spectrum seems to converge at first, but actually diverges for high values of $l_{\rm max}$, which in practice depends on various parameters, with the typical behavior of the asymptotic series. In keeping with this view one can attribute a meaning to the finite result obtained by truncating the MSE: at a given $l_{\rm max}$ the difference between the exact and the approximate result can be made very small, provided the independent parameters of the theory (like the photo-electron energy and the amount of overlap between the bounding spheres) lie in a definite domain of their definition space. For energies in the near edge region this can be obtained by taking a moderate overlap between bounding spheres, of the order of 30$\sim$40\%, so as to reduce the space between them and their MT spheres. Empty spheres can be added to satisfy this condition. 

\begin{figure}

  \begin{center}
    \resizebox{58mm}{!}{\includegraphics{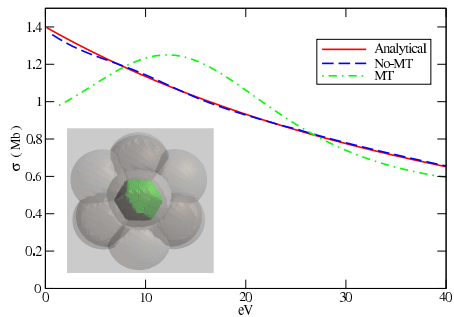}}
    \caption{(color online) Cross section for Li$^{2+}$ with 15 cells compared to the
             analytical result. The solution for a MT central sphere is also
             shown.}
    \label{fig_li2+}
  \end{center}

\end{figure}

We have tested the present FP-MS scheme against the analytical solution of the 
absorption cross section for hydrogen-like atoms given by ~\cite{bethe77}

$$
  \sigma \, ( \, k \, ) = 4 \pi^2 \, \alpha  \frac{ 2^7 }{ 3 }
  \frac{ 1 }{ Z^2 }
  \left( \frac{ 1 }{ 1 + ( \frac{ k }{ Z } )^2 } \right)^4 \,
  \frac{ e^{ -4 \frac{ Z }{ k } \tan^{ -1 } ( \frac{ k }{ Z } ) } }
       { 1 - e^{ -2 \pi \frac{ Z }{ k } } }
$$
in the case of the Li$^{2+}$ atom ($Z=3$). Even though the potential is spherically symmetric in the whole space with respect to the atomic center so that it is easy to reproduce numerically the cross section, this is not obvious in the MS scheme. 
To this purpose we have partitioned the space inside a sphere of radius $R$ = 8.6 au into an atomic sphere of 4.15 au and 14 other empty spheres, all truncated so that the resulting polyhedra do not overlap and such that their bounding sphere do not overlap more than 40\%. To calculate the contribution of the outer sphere we integrated inwardly the Coulomb potential. Fig. \ref{fig_li2+} shows the almost exact agreement between the analytical and the numerical result, indicating that the partitioning procedure for solving the SE is able to reconstruct the global solution. Moreover the oscillations due to the truncation of the potential inside each cell (shown by the solution for a truncated central sphere with radius 4.15 au) cancel each other, showing that at a common boundary the overall solutions inside two adjacent cells are continuously smooth. For this test a value of $l_{\rm max} = 4.15 \, \sqrt{3} \sim 8$ was taken at the end of the energy interval $E_{\rm max} = 3$ Ryd.

\begin{figure}

    \resizebox{88mm}{!}{\includegraphics{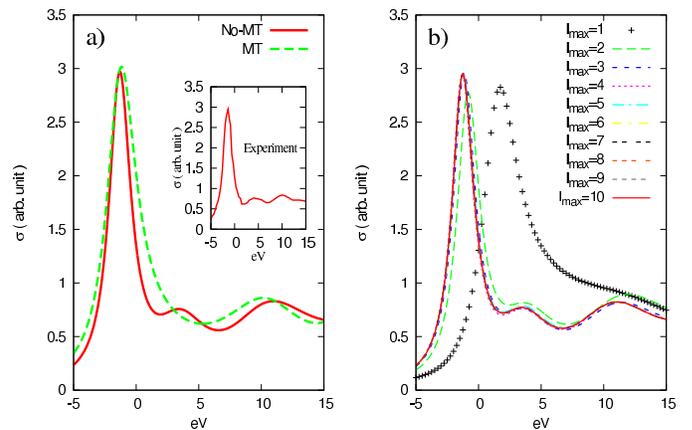}}

    \caption{(color online) (a) Cross section for GeCl$_4$ molecule with nine scattering cells
             located at the sites of a BCC lattice, compared with the MT result and experiment. (b) Study of its convergence rate as a function of $l_{max}$ up to $l_{max} = 10$.}
    \label{fig_gecl4}

\end{figure}

Fig. \ref{fig_gecl4}a shows an application of the method to the calculation of the Ge K-edge absorption spectrum of the tetrahedral molecule GeCl$_4$ ~\cite{filipponi98}. The MT approximation could never reproduce the first bump after the main transition. Its appearance is due to the introduction of the anisotropy of the potential inside the atoms and the presence of four empty Voronoi cells completing the BCC unit cell. An $l_{\rm max} = 4$ was sufficient to reach convergence of the spectrum, as verified by using higher $l$ values up to $l_{\rm max} = 10$ (Fig. \ref{fig_gecl4}b). 





%


%

%
%










%
%

%

%
In conclusions we have developed a FP-MS scheme which is a straightforward 
generalization of the usual theory with MT potentials and implemented 
the code to calculate the cross section for several spectroscopies like 
absorption, photo-electron diffraction and anomalous scattering.
The key point in this approach is the generation of the cell solutions 
$\Phi_{ L }\, (\, { \bf r } \,)$ for a general truncated potential free of the 
well known convergence problems of AM expansion together with an alternative derivation of the MSE which allows us to treat the matrices $S$ and $E$ as square, with only one truncation parameter given by the classical relation $l_{\rm max} \sim k \, R$. Even though this truncation procedure does not converge, taking a moderate overlap between bounding spheres assures satisfactory result in the approximation of the exact solution.
At the same time we have provided an efficient and fast method for solving numerically a partial differential equation of the elliptic type in polar coordinates which can also be used to solve the Poisson equation. 



%
K. Hatada and C.R. Natoli wish to thank Dr. P. Kr\"uger for many
useful discussions.

\bibliographystyle{apsrev}

\bibliography{myrefs04}

\end{document}